%% file: main.tex
\documentclass[runningheads]{llncs}

\usepackage{amssymb,amstext,amsmath}
\usepackage{epsfig,graphicx,times}
\usepackage{pgfplots}
\usepackage[hyphens]{url}
\usepackage{algorithm}
\usepackage{algorithmic}
\usepackage{caption}
\usepackage{multicol,multirow}
\usepackage{booktabs}
\usepackage{rotating}
\usepackage{tabularx}
\usepackage{wrapfig}
\usepackage{subfigure} 

\begin{document}

\title{How to Search the Internet Archive Without Indexing It}

\titlerunning{How to Search the Internet Archive Without Indexing It}

\author{Nattiya Kanhabua\inst{1}, Philipp Kemkes\inst{2}, Wolfgang Nejdl\inst{2}\\Tu Ngoc Nguyen\inst{2}, Felipe Reis\inst{2}, Nam Khanh Tran\inst{2}}
\authorrunning{Kanhabua et al.}

\institute{
	Department of Computer Science, Aalborg University, Denmark \and
	L3S Research Center / Leibniz Universit\"{a}t Hannover, Germany
}

\maketitle

\vspace{-0.2cm}
\begin{abstract}
  Significant parts of cultural heritage are produced on the web
  during the last decades. While easy accessibility to the current web is
  a good baseline, optimal access to the past web faces several
  challenges. This includes dealing with large-scale web archive
  collections and lacking of usage logs that contain
  implicit human feedback most relevant for today's web search.  In
  this paper, we propose an entity-oriented search system to support
  retrieval and analytics on the Internet Archive. We use Bing to 
  retrieve a ranked list of results from the current web. 
  In addition, we link retrieved results to the WayBack Machine; thus allowing
  keyword search on the Internet Archive without processing and
  indexing its raw archived content. Our search system complements existing web
  archive search tools through a user-friendly interface, which comes close to
  the functionalities of modern web search engines (e.g., keyword search,
  query auto-completion and related query suggestion), and provides a great 
  benefit of taking user feedback on the current web into account also
  for web archive search. Through extensive experiments, we conduct
  quantitative and qualitative analyses in order to provide insights
  that enable further research on and practical applications of web
  archives.
\end{abstract}

\input{introduction}

\input{relatedwork}

\input{preliminaries}

\input{approach}

\input{experiment}
\input{discussion}

\small{
	\vspace{-0.3cm}
	\section*{Acknowledgments}
	\vspace{-0.25cm}
	This work was partially funded by the European Commission for the ERC Advanced Grant ALEXANDRIA under the grant number 339233.
}
\bibliographystyle{abbrv}
\vspace{-0.2cm}	
\begin{tiny}
\bibliography{websci2016,cacm2013}
\end{tiny}
\end{document}

%% file: introduction.tex
\vspace{-0.2cm}
\section{Introduction}
\vspace{-0.1cm}
Traditional institutions, e.g., national libraries, keep our cultural heritage and need to be complemented with facilities for preservation and public access to online cultural assets. This is critical given that even for the presumably interesting resources shared through social media like
Twitter were estimated that 27\% of those are lost and not archived
after $2\frac{1}{2}$ years~\cite{SalahEldeen:2012:LMR:2403832.2403850}.  National and
international initiatives have recognized this need and started to
collect and preserve parts of the web. The Internet Archive
has by far the largest web archive collection among the institutions
active in web preservation, where it has collected more than 2.5~Petabyte of web content since 1996.  Another important European initiative is the
Internet Memory Foundation, active in several EU-funded research
projects on web archiving, with a set of smaller crawls for specific
topics, domains and projects. Two important national libraries engaged in web preservation are the British Library and the German National Library, with the aim to preserve their national web content.

Easy access to historical web information becomes more and more
important as the means for accessing and exploring these archives, but
the current facilities are severely
underdeveloped~\cite{Costa:2013:SWA:2487788.2488116,Gomes:2011:SWA:2042536.2042590}. None
of the archive initiatives is able to provide their collections
through an interface, which comes close to the functionalities we see
on today's web search engines. The Wayback Machine\footnote{\url{https://archive.org/web/}} provides the
ability to retrieve and access web pages stored in the Internet
Archive. However, it requires users to represent their information needs by
specifying the URLs of web pages to be retrieved; 
thus demanding a lot of manual effort compared to the current web search engines, such as, Google, Yahoo! and Bing.  Clearly, a simple, yet effective search interface
is needed to retrieve information, which is stored in the Internet
Archive and other web archive collections.

\vspace{0.2cm}
\begin{wrapfigure}{l}{0.4\textwidth}
\includegraphics[width=0.4\columnwidth]{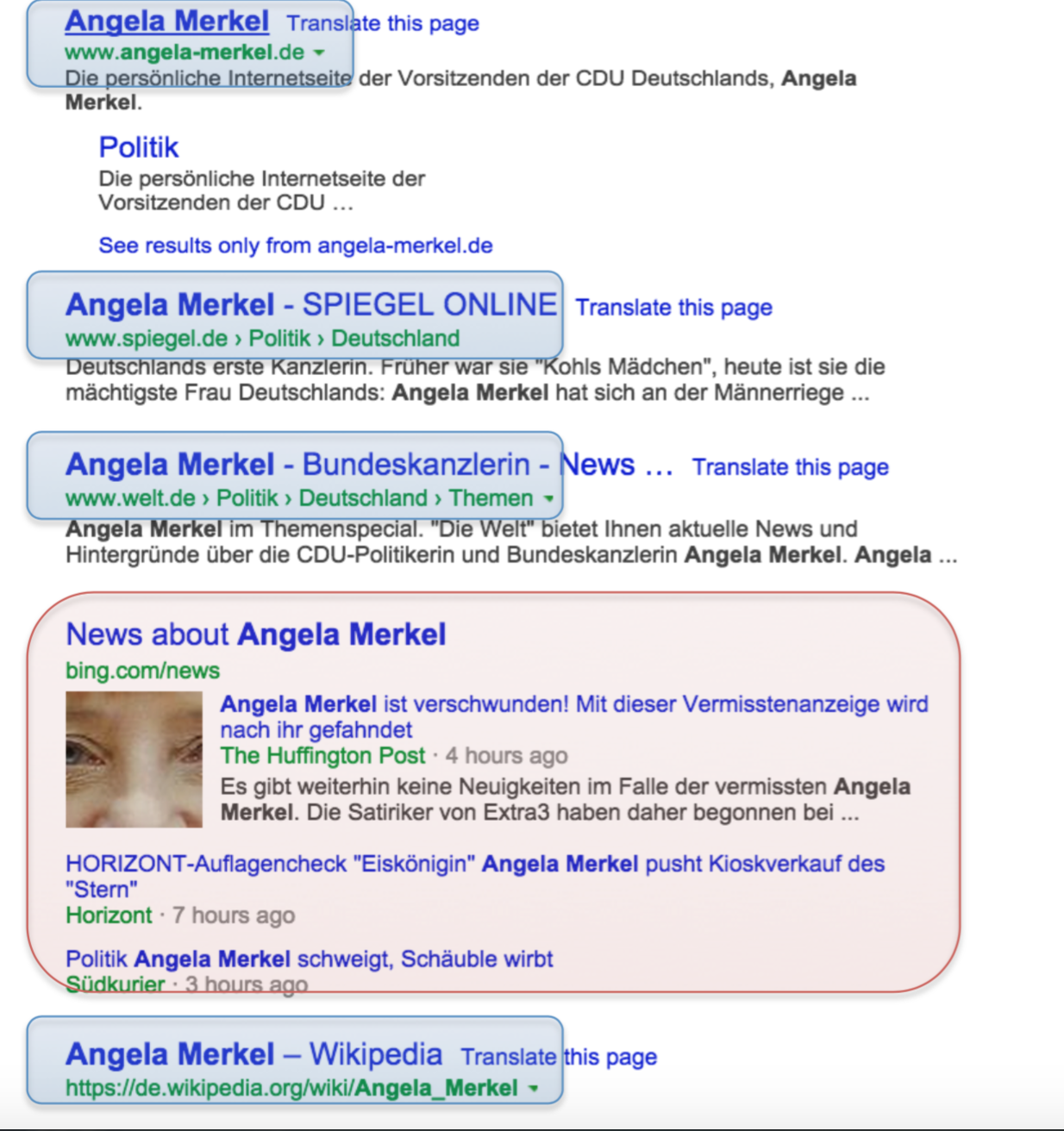}
\caption{URLs in blue areas are \textit{long-term relevant}; URLs in the red area are \textit{short-term relevant}.}
\label{fig:intro}
\end{wrapfigure}

One major problem with web archive search is an absence of query
logs. Without search logs in web archives, it is difficult to understand users' information needs and thus not able to provide
a good ranking of search results without bias. Consider the top search results of the entity \textsf{Angela Merkel} in Fig.~\ref{fig:intro}, which was retrieved from Bing on August 20, 2015. 
Top search results for a popular entity like the German politician consist
of both long-term relevant pages, e.g., the Wikipedia page or
bibliography pages in news websites, and short-term relevant
pages, i.e., news articles about the entity. It can be seen that no news articles aged over one day
appear in the top results, the rest of the search results are long-term
relevant web pages associated with static URLs (those that are unchanged over a long time period).

To compensate this
shortcoming, we built a prototype archive search system on top of
Bing, which already provides a good mix of long-term and short-term
relevant results. Our assumption is that, on the current web search, there are certain types
of query intents that are similar to information needs on web archive
search. In particular, we are interested in supporting web archive
searches for \textit{named entity queries}, which
represent a significant fraction of current web search
queries~\cite{Miliaraki:2015:SGM:2736277.2741284,Yin:2010:BTW:1772690.1772792}.

The goal of this work is to provide a scalable and responsive
search system that supports entity-oriented search on web archives. 
We propose a novel web archive search system that
leverages a current web search engine and the Internet Archive.
Relying on commercial web search engine technologies for accessing
web archives help us to achieve good quality ranking results (with high precision) based on
search sessions and implicit human feedback.  While providing
entity-based indexing of web archives is crucial, we do not address
the indexing issue in this work, but instead extend the WayBack
Machine API in order to retrieve archived content. 

For the best of our knowledge, we are the first
to provide entity-oriented search on the Internet
Archive, as the basis for a new kind of access to web archives, with
the following contributions: (1) We propose a novel web archive search system that supports 
entity-based queries and multilingual search. (2) We make our search system publicly
accessible for enabling further research on and
practical applications for web archives. (3) Through extensive
experiments, we conduct qualitative and quantitative studies and
provide detailed analysis on the results returned through our web
archive search system. (4) Finally, we outline the next steps
towards more advanced retrieval and exploration of web archive
content.

The organization of the rest of the paper is as follows. 
In Section~\ref{sec:relatedwork}, provides a discussion of related work. In Section~\ref{sec:preliminaries}, we present our problem statement. In Section~\ref{sec:approach}, we describe our proposed web archive search and the underlying methodology. In Section~\ref{sec:experiments}, we present the evaluation of our proposed approaches and discuss the experimental results. Finally, we conclude our work in Section~\ref{sec:discussion}.

%% file: relatedwork.tex
\vspace{-0.2cm}	
\section{Related Work}
\label{sec:relatedwork}
\vspace{-0.1cm}
The Internet Archive is
a non-profit organization with the goal of preserving digital document
collections as cultural heritage and making them freely accessible
online. Another important European initiative is the Internet Memory Foundation, active in several EU funded research projects on web archiving. Two important national libraries engaged in web preservation are the British Library and the German National Library, with the aim to preserve national web content. Despite an enormous amount of information is stored in web archives,
there are a few search prototypes to provide access to these archives,
but all come with a number of limitations~\cite{dougherty2009historical}. 
In 2009, the Internet Archive
ran a pilot in providing full-text searchability for parts of their
archive, making the first five years of their web archive (1996-2000)
available for searching. However, the search ranking mechanisms available at
that time were not adequate, and the search results were full
of spam; thus limiting users from advanced search and exploration of archived content.

The Wayback Machine, a web archive
access tool developed by the Internet Archive, provides the ability to retrieve and
access web pages stored in a web archive, but it requires a user to
access data by specifying the URL of a web page to be retrieved. For
example, given the query URL \url{http://www.usa.gov}, search results are displayed in a calendar view showing the number
of times the URL was crawled (not how many
times it was actually updated). To date, it is not possible to search by
keywords. Similarly, the Memento
project\footnote{\url{http://timetravel.mementoweb.org}}
provides access to previous versions of a web page existed at some dates in the past, 
by entering the web page's URL, and by specifying the desired date in a browser plug-in. 
In this manner, Memento makes archived content discoverable via the original URL
that the searcher already knew about, and redirecting the user to
the archive, which hosts the page at the time indicated by the user. 
Archive-IT\footnote{\url{https://archive-it.org}} is a web archiving
service for collecting and accessing cultural heritage sites on the
web, built by the Internet Archive. The service supports organizations
to harvest, build, and preserve collections of digital content, as well as full-text search on the archived collections.
Nevertheless, this search functionality is only limited to a set of smaller crawls for specific topics, domains and projects.
 
In the context of searching web archives, Nguyen et al.~\cite{Nguyen:2015:TRW:2766462.2767832} 
proposed an approach to discovering important
documents along the time-span of the web archives by combining relevance, temporal authority,
diversity and time in a unified ranking framework.
Singh et al.~\cite{Singh:2016:HDH:2854946.2854959} proposed a novel algorithm, HistDiv, 
that explicitly models the aspects and important time windows for supporting historical search intent. 
For an application of web archives, Tran et al.~\cite{Tran:2015:BNS:2806416.2806486} studied a timeline summarization of an entity served as important memory cues in a retrospective event exploration.

%% file: preliminaries.tex
\vspace{-0.2cm}
\section{Problem Statement}
\label{sec:preliminaries}
\vspace{-0.1cm}

\textbf{Information needs.} Entity queries, e.g., person, organization and location,
comprise a significant fraction in web search logs~\cite{Miliaraki:2015:SGM:2736277.2741284,Yin:2010:BTW:1772690.1772792}. Due
to a lack of web archive search logs, we assume similar search behavior, i.e., information needs are either
(1)~exploring entity-related information, or (2)~seeking a specific event in which entities involve. We allow users to search archives using entity queries that exist in
Wikipedia for both English and German. To enhance query formulation, 
we provide two techniques, namely, auto-completion and related-entity suggestion. Finally, we also determine the categories for a given entity, by using a heuristic approach~\cite{Kanhabua:2010:ETS:1816123.1816135} for a further analysis.

\textbf{Search results.} In our context, we define two
main types of relevant results for entity queries: (1)~general or static
pages about the entity that do not change much over time (which are
relevant over a long period of time), so-called \textit{long-term
  relevant} and (2)~dynamic pages such as news articles or blogs
(which are only relevant for a short-time period), denoted
\textit{short-term relevant}. 

\textbf{Ranking method.} In this work, we aim at providing high precision search results, 
rather than optimizing recall.  In order to
achieve our goal, we build on the ranking of search results provided
by Bing, given that the current web search engines already try to
provide a suitable mix of long-term and short-term relevant pages,
while taking a lot of user feedback into account. Although we assume
Bing returns relevant results, we still need to investigate search
results for different kinds of entities in a principled way.

\textbf{Result coverage.} Coverage of 
results returned by Bing is concerned about how many of these results
are archived on the Internet Archive.  We hypothesize that many of
these general pages about the entity are archived already
, while news or recent pages are not be indexed
yet. 
In fact, important news sites or domains themselves
should all be archived. Another aspect to be addressed is the
\textit{temporal dynamics} of search results. We assume that the
top-ranked results returned by Bing at different time points and rate
of their changes vary by entity categories. Finally, we will 
analyze result variations over time in order to gain more insight.


%% file: approach.tex
\vspace{-0.2cm}	
\section{Our Approach}
\label{sec:approach}
\vspace{-0.1cm}	

\begin{figure}[t!]
\hspace{-0.3cm}
	\mbox{\subfigure[]{\includegraphics[width=0.545\columnwidth]{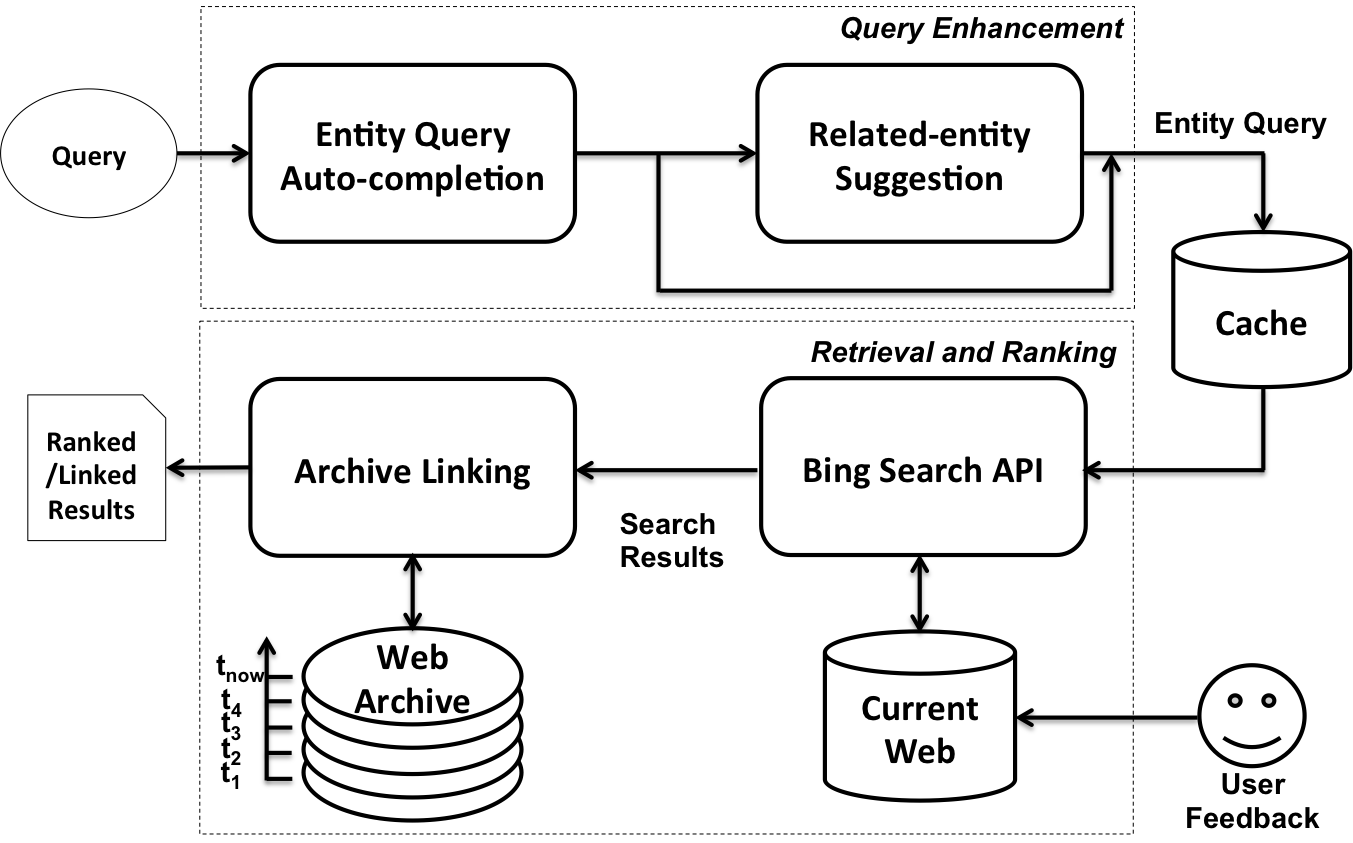}}\quad
		\subfigure[]{\includegraphics[width=0.45\columnwidth]{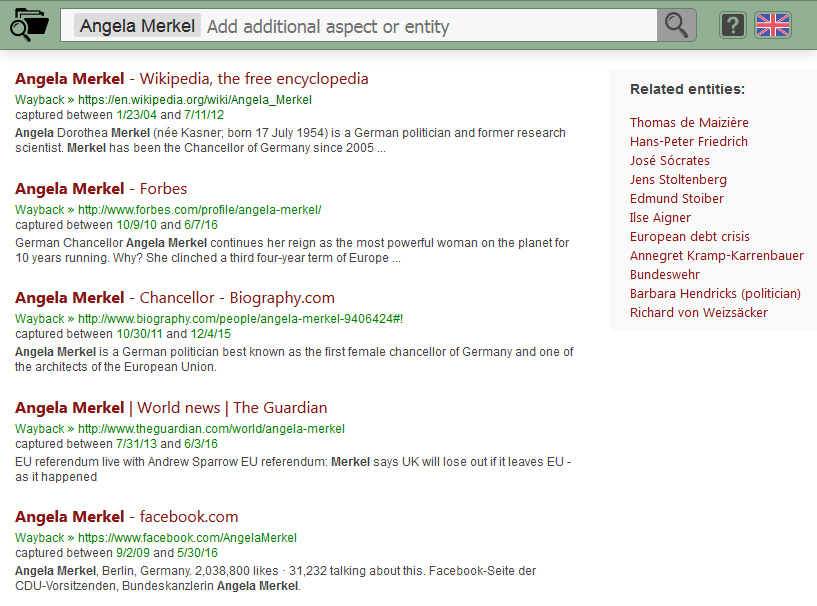}}}
	\caption{Entity-oriented web archive search: (a) system framework and (b) user interface.}
	\label{fig:system}
\end{figure} 
The huge size of web archives has not only created great challenges for
indexing them, but also increased the difficulty for users to express
their information needs. More precisely, it is not easy for users
to compose a succinct and precise query because of the temporal
dimension.  Our archive search system provides keyword-based search
functionalities, similar to existing web search engines.  Users can
issue an entity-based query for any entity described in Wikipedia in
English and German. The system returns a ranked list of search
results, which provides links to both the current page on the web, as
well as the archived versions in the Internet Archive, using blue and
green links. Blue link refers to the current page on the web. Green
link refers to the archived versions in the Internet Archive.  Our
system is publicly accessible\footnote{\url{http://alexandria-project.eu/archivesearch/}}. Fig.~\ref{fig:system}
shows the overview of our system framework and a search user interface. In the
following, we will describe the proposed approach underlying different components, which comprise query auto-completion, Bing search API,
archive linking, caching, and related entity suggestion.
  
\subsection{Query Auto-Completion}
\label{subsec:autocomplete}
\vspace{-0.2cm}	
\input{autocompletion}

\subsection{Bing Search API, Archive Linking and Result Caching}
\vspace{-0.2cm}	
Bing is Microsoft's search engine providing access to their current
web index through a RESTful API available at the Azure
Marketplace. Bing returns results in XML or JSON data formats and
offers two different API endpoints: (1) the full featured \textit{Bing
  Search API}, and (2) the restricted and less expensive \textit{Bing
  Search API - web Results Only}. The later lacks of few meta data
like the overall result count. Nevertheless, it provides all basic
search result information, such as URL, title and a text snippet. By
specifying parameters, we can request optimized results for different
languages/countries.  We therefore use the endpoint with web results
only. Yahoo!  and Google provide similar search APIs but at higher
costs with more restriction.

After obtaining search results from the Bing search API, we link the
ranked list of results to the WayBack Machine to support browsing
through the archived versions of web pages. The WayBack Machine is a
tool provided by the Internet Archive that allows access to its web
archives by specifying a URL. The URL-based access can be
programmatically used through an API provided by the Internet
Archive. For a given URL it returns a list of all dates when the URL
has been archived. When a URL has been archived many times in the
past, retrieval can take very long time. Therefore, we use two requests
to retrieve only the first and last capture dates to display the
time span at which the web pages has been archived. When the temporal intent
of the user is provided, we narrow down to return only
the 
revisions around the interested time point.

To avoid recurring requests to the Bing and WayBack Machine APIs, we
store the search results locally in our cache, using a simple
relational database. In order to take into account the fact that
search results change over time, we update search results monthly to
keep our cache up-to-date and to track changes at both sources.
Besides the queries entered by our users, we use also the 10,000 most
viewed English Wikipedia entities as queries. As a side effect, this
procedure results in building a corpus of past search results, which
will support promising, longitudinal studies of web archive search,
investigating how results change over time, triggered by events or
changing user behavior. In Section~\ref{sec:experiments}, we will
analyze the cache in order to reveal several important aspects, e.g.,
how long a web page stays relevant and when it fades away from
top-ranked results. This insight will help improving the next
version of our web archive search prototype.

\subsection{Related Entity Suggestion}
\label{subsec:suggestion}
\vspace{-0.2cm}

\input{querysuggestion} A traditional web search engine supports
exploration by suggesting related queries, which is based on analyzing
search sessions and identifying the co-occurrences of the issued
query. For web archive search, we do not have query logs for obtaining
search sessions, thus we leverage a dump of Wikipedia articles and
build an entity graph in order to find related queries for our
entity-oriented search. We follow the approach to determining the
link-based entity relatedness originally proposed
in~\cite{milne2008learning}. Relatedness between two entities $e_{1}$ and $e_{2}$ is measured based
on the overlap between the set of Wikipedia articles that link to
$e_{1}$ and the set of articles that link to $e_{2}$.

\begin{equation}
relatedness(e_{1}, e_{2}) = \frac{log(max(|S_{1}|,|S_{2}|)) - log (|S_{1} \cap S_{2}|)}{ log (|W|) - log(min(|S_{1}|,|S_{2}|))}
\end{equation}

\noindent where $S_{i}$ is the set of articles that links to entity $e_{i}$, and $W$ is the set of all articles in Wikipedia. Our entity graph is constructed from a recent Wikipedia dump
(downloaded from September 2015 for both English and German
Wikipedia), with the assumption that the link-based relationships
between a pair of two entities can be accumulated; thus relaxing time
sensitivity (in this case, the relatedness does not bias to any time
point). An interesting extension will be to take time dependent
relationships into account, which is planned for the next version of our
system.

\subsection{Multilingualisms}
\label{subsec:multilingual}
\vspace{-0.2cm}	
The described components are designed to support different
languages. Dependent on the language selected by the user, we use the
corresponding Wikipedia version to generate the query auto-completion
and the related entities suggestions. More importantly, we request
search results from Bing, which are optimized for the selected language
and region. Furthermore, we leverage Wikipedia inter-language links
when a user changes the front-end language.  For example: When an
English user searched the term ``climate change'' and switches to
German, he will be redirected to ``Klimawandel''. Currently, we
support English and German, but will add additional languages in the
future.


%% file: autocompletion.tex
We support the query formulation process using query
auto-completion. When a user types a query, we suggest a short list
of relevant entities in order to help the user complete his/her
information needs. We use a Wikipedia
entity index comprised of all Wikipedia entities and store it in a
trie data structure to allow fast prefix lookup. Additionally, we
split all entities at white and special characters. All strings
starting at each token are added to the trie as an additional reference
to the original entity. Furthermore, we also take into account
simplified versions of all tokens which contained letters with accents
in our index.  This allows our application to suggest entities even if
the user does not know the exact name or cannot type the name in a
foreign alphabet. As an example, for the query ``schroder'' we would
suggest the former German chancellor ``Gerhard Schr{\"o}der''. We
further rank the suggested query completions by their popularity using
the cumulative page views (see the detailed description below). To
penalize the time-sensitive popularity of the entities, the daily
page views are accumulated over a long period. Finally, the entity
selected by the user is sent as input to the search API.

%

%
%

Wikipedia page
views\footnote{\url{https://en.wikipedia.org/wiki/Wikipedia:Pageview_statistics}}
are statistics consisting of the number of times a particular Wikipedia page has been
requested over time. 
In Wikipedia, the view counts for pages that redirect to a given page
are not combined with page views of the page being redirected
to. 
In this work, we aggregate all these related views to present the
popularity (reflected by the page views) of an entity query for all
its query variants. We computed the aggregated statistics of page
views approximately a period of 4 years (2011 to 2015).

%
%

%% file: experiment.tex
\vspace{-0.2cm}	
\section{Experiments}
\label{sec:experiments}
\vspace{-0.1cm}	
We conducted extensive experiments in order to gain insight into our assumptions presented in Section~\ref{sec:preliminaries}. We seek to answer two main research questions as follows.\\

\noindent \textit{RQ1. What is the coverage of archived content retrieved by the current search engine?}\\
\vspace{-0.2cm}	

\noindent \textit{RQ2. To what extent the search results change over time, and why they change?}\\
\vspace{-0.2cm}	

In the following, we will divide our experimental results into two
main parts, where we describe our quantitative and qualitative
analyses for each of the aforementioned research questions.

\input{experiment_q1_overall}

\input{experiment_q1_examples}

\input{experiment_q2_overall}

\input{experiment_q2_examples}

%% file: experiment_q1_overall.tex
\begin{figure}[t!]
	\centering
	\includegraphics[width=0.75\columnwidth]{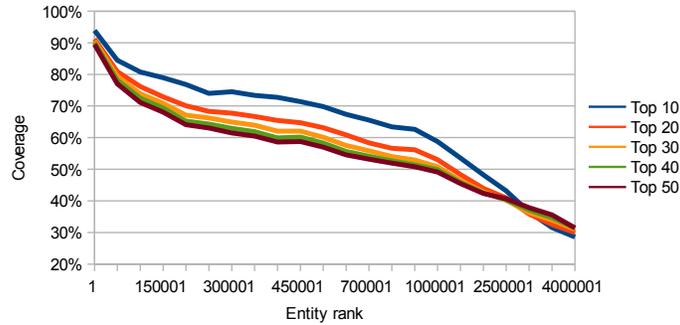}
	\caption{Coverage (percentage) of archived content at different top-k results over entities ranked by their popularity.}
	\label{fig:content_retrievability}
\end{figure}

\textbf{Part I: Analysis Results for RQ1.} Our system relies on the assumption that many pages returned as search results for entity queries are archived by the Internet Archive. To check this assumption, we took all English Wikipedia entities sorted by their view count and selected buckets of 1,000 entities at different positions in this list to represent entities from different popularity categories. We started with the 1,000 most viewed entities and continued with the entities from position 50,001 to 51,000 and so on. For each entity in an individual bucket, a search query was conducted and we checked how many results on the first five pages (10 results per page) were archived by the Internet Archive. Table~\ref{tab:content_retrievability} and Fig.~\ref{fig:content_retrievability} show the average results per page and bucket.

The results of popular entities (rank: 1 - 1000) show a very high
coverage with the Internet Archive. On page one, 94\% of the results
are archived. On pages two to five, still 87\% to 89\% are available
at the Internet Archive. Overall the coverage declines for less
popular entities. Interestingly, it drops faster for the first pages
of the search results than for the posterior ones. Upon inspection,
this seems to be caused by the fact that Bing ranks recent results
higher (for example on the first page of its search results), while
the Internet Archive needs much more time to archive less popular
pages.

\begin{table}
\centering
{\scriptsize
\setlength\tabcolsep{3.5pt}
	\begin{tabular}{c || c | c | c | c | c | }
Entity rank       & Top 10 & Top 20 & Top 30 & Top 40 & Top 50 \\ \hline
~~~~~~1 - 1000          & 94\%   & 91\%   & 91\%   & 90\%   & 89\%   \\
50001 - 51000     & 85\%   & 81\%   & 79\%   & 78\%   & 77\%   \\
100001 - 101000   & 81\%   & 76\%   & 74\%   & 72\%   & 71\%   \\
150001 - 151000   & 79\%   & 73\%   & 71\%   & 69\%   & 68\%   \\
200001 - 201000   & 77\%   & 70\%   & 67\%   & 65\%   & 64\%   \\
250001 - 251000   & 74\%   & 68\%   & 66\%   & 64\%   & 63\%   \\
300001 - 301000   & 74\%   & 68\%   & 65\%   & 63\%   & 61\%   \\
350001 - 351000   & 73\%   & 67\%   & 64\%   & 62\%   & 60\%   \\
400001 - 401000   & 73\%   & 65\%   & 62\%   & 60\%   & 59\%   \\
450001 - 451000   & 71\%   & 65\%   & 62\%   & 60\%   & 59\%   \\
500001 - 501000   & 70\%   & 63\%   & 60\%   & 58\%   & 57\%   \\
600001 - 601000   & 67\%   & 61\%   & 58\%   & 56\%   & 55\%   \\
700001 - 701000   & 66\%   & 58\%   & 56\%   & 54\%   & 53\%   \\
800001 - 801000   & 63\%   & 57\%   & 54\%   & 53\%   & 52\%   \\
900001 - 901000   & 63\%   & 56\%   & 53\%   & 51\%   & 51\%   \\
1000001 - 1001000 & 59\%   & 53\%   & 51\%   & 50\%   & 49\%   \\
1500001 - 1501000 & 54\%   & 48\%   & 46\%   & 46\%   & 45\%   \\
2000001 - 2001000 & 48\%   & 44\%   & 43\%   & 43\%   & 42\%   \\
2500001 - 2501000 & 43\%   & 41\%   & 40\%   & 40\%   & 41\%   \\
3000001 - 3001000 & 36\%   & 36\%   & 36\%   & 37\%   & 38\%   \\
3500001 - 3501000 & 32\%   & 33\%   & 34\%   & 35\%   & 36\%   \\
4000001 - 4001000 & 29\%   & 30\%   & 31\%   & 31\%   & 31\%  
	\end{tabular}
\vspace{0.2cm}
	\caption{Coverage (percentage) of archived content at different top-k results over entities ranked by their popularity.}
	\label{tab:content_retrievability}
	\vspace{-0.1cm}
}
\end{table}


\begin{figure}[t!]
	\centering
	\includegraphics[width=1\columnwidth]{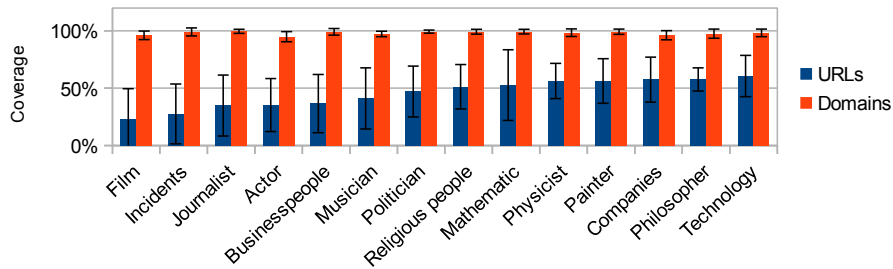}
	\caption{Coverage (percentage) of archived URLs and domains at top-100 results for different entity categories.}
	\label{fig:coverage_by_category}
\end{figure}

To gain more insight, we conducted a coverage study by entity
categories. More precisely, we analyzed top-100 search results for 300
popular entities from 14 different categories, for example, actor,
journalist, painter, and politician, where there are approximately
20 entities per category. In this study, we only considered search results with .DE
domains (German web pages), and checked the coverage with our local
German web archive
, instead of web
archives of the Internet Archive. As shown in
Fig.~\ref{fig:coverage_by_category}, the coverage statistics on the
German web archive shows significantly lower results than the one
based on comparison with the Internet Archive due to search result
bias towards English web pages, in general, even for the German
version of Bing. It can be observed that categories with lower coverage tend to associate with
recent and dynamic web content, whereas the results of the categories
with higher coverage are rather static and less changed. Note that, 
result URLs are not always archived (e.g., for newspaper articles), but
nearly all domains (i.e., news sites) are archived, regardless of
entity categories.

We also performed search result annotation of 9 entities that were manually selected. We
employed 5 human annotators to label top-100 search .DE
results (by filtering out non .DE domains). For each (query, URL) pair, we
asked at least 4 assessors to give a label based on relevance
assessment criteria consisting of three scales: \textit{long-term
  relevant}, \textit{short-term relevant} and
\text{unknown}. 
The results are shown in Table~\ref{tab:annotation}, where we can
notice that non-active entities, such as, Pablo Picasso and Ernest
Hemingway, have more long-term relevant results than active entities
like Elon Musk and Leonard Nimoy (to be expected).

\begin{table}[htbp]
	\centering
		{\scriptsize
	\begin{tabular}{lrrr}
		\toprule
		\textbf{Entity} & \textbf{Category} & \textbf{Long-term} & \textbf{Short-term} \\
		\midrule
		Leonard\_Nimoy & Actor & \multicolumn{1}{c}{52.00\%} & \multicolumn{1}{c}{48.00\%} \\
		Elon\_Musk & Business people & \multicolumn{1}{c}{37.50\%} & \multicolumn{1}{c}{62.50\%} \\
		Costa\_Concordia\_disaster & Incidents & \multicolumn{1}{c}{52.40\%} & \multicolumn{1}{c}{47.60\%} \\
		Ernest\_Hemingway & Journalist & \multicolumn{1}{c}{81.48\%} & \multicolumn{1}{c}{18.52\%} \\
		Giuliana\_Rancic & Journalist & \multicolumn{1}{c}{40.00\%} & \multicolumn{1}{c}{60.00\%} \\
		Pablo\_Picasso & Painter & \multicolumn{1}{c}{97.60\%} & \multicolumn{1}{c}{2.40\%} \\
		Banksy & Painter & \multicolumn{1}{c}{48.10\%} & \multicolumn{1}{c}{51.20\%} \\
		Vietnam & Politics & \multicolumn{1}{c}{100.00\%} & \multicolumn{1}{c}{0\%} \\
		Ku\_Klux\_Klan & Politics & \multicolumn{1}{c}{12.50\%} & \multicolumn{1}{c}{87.50\%} \\
		\bottomrule
	\end{tabular}%
	\caption{Percentage of long-term relevant and short-term relevant pages in top-100 results (filtered out non .DE domains).}
		\label{tab:annotation}%
}
\end{table}

%% file: experiment_q1_examples.tex
In the following, we provide analyses for two selected entities in
order to better understand the coverage aspect of search results and
web archives.

\paragraph{Lady Gaga (view count rank: 108; views: 9,453,966; querying date: 19.01.2016)}

\begin{wrapfigure}{l}{0.1\textwidth}
	\includegraphics[width=0.9\linewidth]{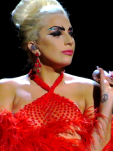} 
\end{wrapfigure}

Lady Gaga is a famous American artist. Her Bing top-50
results have almost complete coverage on the Internet Archive. More
specifically, 98\% of the results are archived. Among the results,
only one URL\footnote{\url{
    http://www.nydailynews.com/entertainment/gossip/linda-perry-slams-lady-gaga-article-1.2500319}}
is not archived. This URL points to a gossip news article inside the
entertainment section of the New York Daily News web site that only mentions the entity in question.
As the entity is not the core mention, the article has relatively low relevance. 
We checked again on January 25, 2016, and the URL was
still not archived. The URL was published online on January 18, 2016. Thus, we interpret that low relevance
news are unlikely to be archived.

\paragraph{Battle of Rathmines (view count rank: 1,000,798; views: 8,723; querying date: 18.01.2016)}

\begin{wrapfigure}{l}{0.1\textwidth}
	\includegraphics[width=0.9\linewidth]{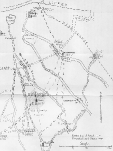} 
\end{wrapfigure}
The Battle of Rathmines was fought in the area what is now the
Dublin suburb of Rathmines in August 1649, during the Irish
Confederate Wars, the Irish theatre of the Wars of the Three
Kingdoms. 
The number of page views for this entity is 8,723, which is not very popular. Nevertheless, about 56\% of the top-50 Bing
results are archived. Among the other 44\% of results, which are not archived,
is the (4th ranked) page\footnote{\url{http://www.thefullwiki.org/Battle_of_Rathmines}}
from a website that shows information from Wikipedia, and the (48th ranked) page\footnote{\url{http://irelandinhistory.blogspot.de/2014/08/blog-post_11.html}} that is a blog post. This entity is related to a real event in Ireland history
that took place in almost 500 years ago, but is very local and rather
unimportant outside of Ireland.

%% file: experiment_q2_overall.tex
\textbf{Part II: Analysis Results for RQ2.} In this section, we present the analysis of Bing results for
entity-based queries executed in three time periods: June 2015, August
2015 and January 2016. For a given entity, we computed the overlap
between the top-ranked results at different time periods. Through
this, we gain insight into our question RQ2, on how Bing query
results change over time. We conducted a study on 300 popular queries
of 14 different categories as explained in RQ1. We discuss a few
samples ranging from low to high overlapping rates. 

\begin{figure}[t]
	\centering
	\includegraphics[width=1\columnwidth]{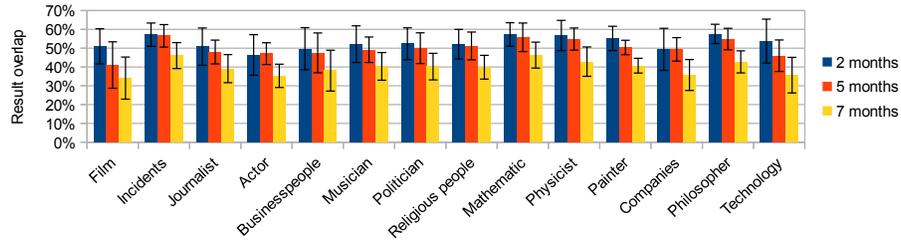}
	\caption{Result overlap for different entity categories.}
	\label{fig:overlap1}
	\vspace{-0.1cm}
\end{figure}

\begin{figure}[t]
	\centering
	\includegraphics[width=0.7
	\columnwidth]{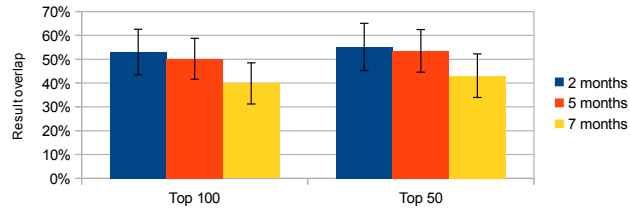}
	\caption{Result overlap for top-50 and top-100 results.}
	\label{fig:overlap2}
	\vspace{-0.1cm}
\end{figure}

Fig.~\ref{fig:overlap1} illustrates the change of search results (as
measured by the overlap statistics) over time for different entity
categories over the period of 2, 5 and 7 months, respectively. In
general, result change after 2 months results in approximately $47\%$
of the top-100 URLs not returned any more. After 7 months, result
change increases to $60\%$. Across different categories, there is not
much difference in temporal dynamics for the search results. The
\textit{Actor} category varies most, whereas the result variation is
least for \textit{Philosopher}. The main explanation here is that the
entities in our \textit{Actor} category are more active than entities
in the \textit{Philosopher} category, thus having more short-term
relevant web pages in the top-100 result list, which change over
time. We observe in Fig.~\ref{fig:overlap2} that the overlap in the
top-50 results is slightly higher than in the top-100 results. This
indicates there is less result change in the first 50 results than
in the next 50 results. However, the difference is not significant.

%% file: experiment_q2_examples.tex
\paragraph{Entity: Barack Obama}

\begin{wrapfigure}{l}{0.1\textwidth}
	\includegraphics[width=0.9\linewidth]{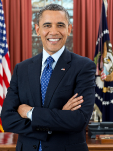} 
\end{wrapfigure}
For the President of the United States 47.5\% of the results from August overlap
with the results retrieved in June. These shared results are URLs
pointing to biography or permanent pages, and most of them appear in
high ranks. The remaining 52.5\% are URLs mostly to news or
categories inside news web portals, which change over time.
The overlapping of search results in this context decreases proportional
to the number of recent events related to the entity: the higher the
proportion of news results, the lower the overall overlapping with
future result sets.

\paragraph{Entity: Donald Trump}

\begin{wrapfigure}{l}{0.1\textwidth}
	\includegraphics[width=0.9\linewidth]{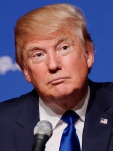} 
\end{wrapfigure}
For Donald Trump, more than 80\% of the URLs are news (National Journal,
newsobserver.com, NBC). Most of these news articles published in June and
do not overlap in August and January. Our data analysis in this experiment
shows that the URLs from the news category, which is listed in results
from June will probably not be shown in August or January. Total
overlap average is 24\%, which is very low. Even for the shorter
periods, June to August, and August to January, a result overlap is low with
30\%. Search results for Trump are a very clear sample for fast
changing results, caused by frequent news articles.

%% file: discussion.tex
\vspace{-0.2cm}	
\section{Discussion and Conclusion}
\label{sec:discussion}
\vspace{-0.2cm}	
Although the system described in this paper already provides
interesting functionalities, it is obviously still work in
progress. As one important extension of functionality, we are working
on more complex types of entity-based queries in order to support
exploratory search, e.g., giving a main entity \textit{Donald Trump}
and related search intents. The search intent can consist of an
entity, such as, \textit{Hillary Clinton} aiming to find all events, 
which involve these two entities, and a specific time period such as
\textit{2015-2016} narrowing down search results to a specific time
period, or any contextual query, such as a concept
\textit{presidential campaign}.  

Another important aspect for our future development is to advance our
ranking. As our current method is relying on the Bing search API, we
sacrifice recall.  Learning from Bing over time as well as from our
user logs, we will be able to provide more sophisticated ranking
taking different features into account. Bing results can act as `soft'
ground truth for learning the
model. 
Our ranking model will then be able to return relevant documents, which
are not longer available on the current Web.

Finally, we also work on improving our suggestion components, i.e.,
related entity suggestion to deal with queries having time as another
aspect (e.g., Obama 2008). In the current system, we exploited a
state-of-the-art method to suggest related entities to the query
entity, with the assumption that their relationship strengths are
accumulated over time. This relationship measure is reasonable to
serve for queries with arbitrary relevant time. However, in reality
relationships between entities do change over time, typically
triggered by events. We can therefore return different related
entities for different time periods to the input entity. For instance,
the entities mostly related to \textit{Hillary Clinton} in 2008 should
differ from those in 2012, because of her different political
positions. Moreover, with an exploratory query for example
\textit{Donald Trump} and search intent \textit{Hillary Clinton}, it
is more helpful to recommend the entities which are related to both
\textit{Donald Trump} and \textit{Hillary Clinton}. 

To summarize, we discussed a web archive search prototype that for the
first time supports entity-oriented queries on the Internet
Archive. Our system leverages Bing and the WayBack Machine to allow
users to search the past Web. We provided search functionalities
including keyword search, query auto-completion, query suggestion, and
a ranked list of results, which are close to the current search engine
systems. We conducted extensive analyses that shed light on web
archive search, and included a discussion of future work as well as
ideas/challenges for the next steps.
